\documentclass{article}

\usepackage{graphics}
\usepackage{amssymb}
\usepackage{amsmath}
\usepackage{lastpage}
\usepackage{xcolor}
\usepackage{graphicx}   
\usepackage{hyperref}
\usepackage{float}
\usepackage{setspace}
\usepackage{authblk}
\usepackage{ccaption}% http://ctan.org/pkg/ccaption

\definecolor{mypink}{RGB}{220, 38, 127}
\definecolor{myblue}{RGB}{0, 114, 178}
\definecolor{liteblue}{RGB}{86, 180, 233}

\usepackage[scaled]{helvet}
\usepackage[T1]{fontenc} % Optional:  Use T1 encoding for better font support

\title{Practical indistinguishability in a gene regulatory network inference problem, a case study}

\author[1,*]{Cody E. FitzGerald}
\author[2,3]{Shelley Reich}
\author[1]{Victor Agaba}
\author[1]{Arjun Mathur}
\author[3]{Michael S. Werner}
\author[1,4]{Niall M. Mangan}
\affil[1]{Department of Engineering Sciences and Applied Mathematics, Northwestern University, Evanston, IL 60208, USA}
\affil[2]{Department of Biological Sciences, Walla Walla University, College Place, WA 99324, USA}
\affil[3]{School of Biological Sciences, University of Utah, Salt Lake City, Utah 84112, USA}
\affil[4]{NSF-Simons National Institute for Theory and Mathematics in Biology, Chicago, IL  60611, USA}
\affil[*]{cody.fitzgerald@northwestern.edu}

\date{\today}

\begin{document}

\maketitle

% Computationally inferring mechanistic insights and underlying control structures from typical biological data is a challenging pursuit.

\section{Abstract} 
Determining mechanistic models of gene regulation, especially underlying phenotypic variation, is a central goal of both mathematical biology and modern evolutionary biology. Progress has been made by analyzing high-throughput sequencing data, yet several challenges, involving both common characteristics of experimental data and the model development process, remain that limit the discovery of general principles. Even the highest-quality experimental data come with challenges. There are always sources of noise, a limit to how often we can measure the system in time, and it is impossible to measure all the relevant states that participate in the full underlying complexity. Additionally, there are usually sources of uncertainty in the underlying biological mechanisms, which give rise to multiple competing model structures. To underscore the need for further analysis of structural uncertainty in modeling, we use a meta-analysis across six journals covering mathematical biology and show that a huge number of mathematical models for biological systems are developed each year, but model selection and comparison across model structures appear to be less common. We then walk through a case study involving inference of regulatory network structure involved in a developmental decision in the nematode, \textit{Pristonchus pacificus}, which exhibits an evolutionary novelty of mouth-dimorphism (predator vs. bacterivore). We first perform a careful study of the \textit{practical indistinguishability} of a model structure, or the ability to uniquely infer the structure given the data, across a wide range of synthetic data regimes by refitting both the true model structure and several misspecified models. We then fit 13,824 distinct regulatory network models to gene expression data from three experimental conditions to determine which regulatory features are supported by the data. We discover \textit{model sets}, or collections of models with shared regulatory network features that best fit the data, for each of the three experiments we considered and identify a regulatory network in the intersection of the three model sets. This model describes the data across the experimental conditions and exhibits a high degree of positive regulation and interconnectivity between the key regulators, \textit{eud-1}, \textit{sult-1}, and \textit{nhr-40}. While the biological results are specific to the molecular biology of development in \textit{Pristonchus pacificus}, the comparative modeling framework introduced here can be applied to other systems of gene regulation in an evolutionary developmental (evo-devo) context.

\section{Introduction}

Developing and parameterizing mathematical models from experimental data, can be a powerful tool to explain, understand, and make predictions about how the underlying system operates. However, parameterizing mathematical models through numerical optimization is a challenging endeavor in its own right, due to \textit{properties of biological data} and \textit{structural uncertainty}. Biological experiments across fields often result in data that include sources of intrinsic and extrinsic noise, are sampled infrequently in time, and usually, not all relevant components of the underlying system can be measured directly, even with the most cutting-edge experimental equipment and designs. As a result, we might have well-founded descriptions of some components of the biological system in a model, but multiple mechanistic hypotheses for the less resolved or well-understood components. This leads to multiple competing model structures, and is known as \textit{structural uncertainty}. For example, it may be known that a single gene embedded within a complex network is regulated by a transcription factor, but unclear if the gene experiences activation or repression by the transcription factor. Structural uncertainty is widespread in biological modeling, and subtle changes in the model structure can influence the results. This topic is rarely explored in depth due to the computational burden of distinguishing between model structures in a data-driven manner. Distinguishing between multiple model structures requires navigating high-dimensional cost function landscapes for each model--a process that can be challenging for even a single model, to determine the best fit of each model structure to the data through numerical optimization. 

Fitting even a single model to data through numerical optimization can be challenging for several reasons, including global and location geometry of the cost function landscape associated with models of biological systems and both structural and practical identifiability issues, which can emerge based on properties of typical biological data. Broadly, cost function landscapes for biological systems are known to feature long, flat canyons, where parameter values can widely vary but the model behavior stays qualitatively the same \cite{gutenkunst2007universally}. These landscapes are hard to fully numerically explore, even when using synthetic data that is noise-free. In the context of realistic experimental data, the cost function landscapes can easily become rippled or distorted, making it more difficult to perform parameter estimation. The properties of experimental data can also impact both structural and practical identifiability of the model. Typically, not all relevant states of the system can be measured, leading to hidden state variables. Hidden states can lead to structural identifiability issues, which occur when the the parameters of the model cannot be uniquely inferred based on the structure of the model alone. Both low temporal data sampling and sources of noise can lead to practical identifiability issues, which occur when the parameters of the model cannot be uniquely determined \textit{given the data}. Both structural and practical identifiability issues lead to a lack of uniqueness in the estimated parameters, which can impact the ability to use the model for explanation or prediction. See \cite{preston2025think} for a recent review of identifiability analysis of linear and nonlinear models and the associated computational considerations. Further complications emerge when we consider multiple model structures to take into account structural uncertainty.  

%this is where to split into two paragraphs

Every change to the structure of the model impacts the geometry of the cost function landscape, potentially altering the ease of parameter estimation. Adding additional parameters to a model increases the dimensionality of the cost function landscape, whereas imposing changes to the functional form of the model that keep the number of parameters fixed, for example, the difference between a Hill Function that describes activation or repression of a gene transcription, can distort or bend the geometry of the cost function landscape, so it is important to consider the sensitivity of the modeling results as a function of all possible model structures that emanate from structural uncertainty, especially in our current scientific moment, when so many mathematical models of biological systems are being produced. 
 
\begin{figure}[H]
    \centering
    \includegraphics[scale=0.3]{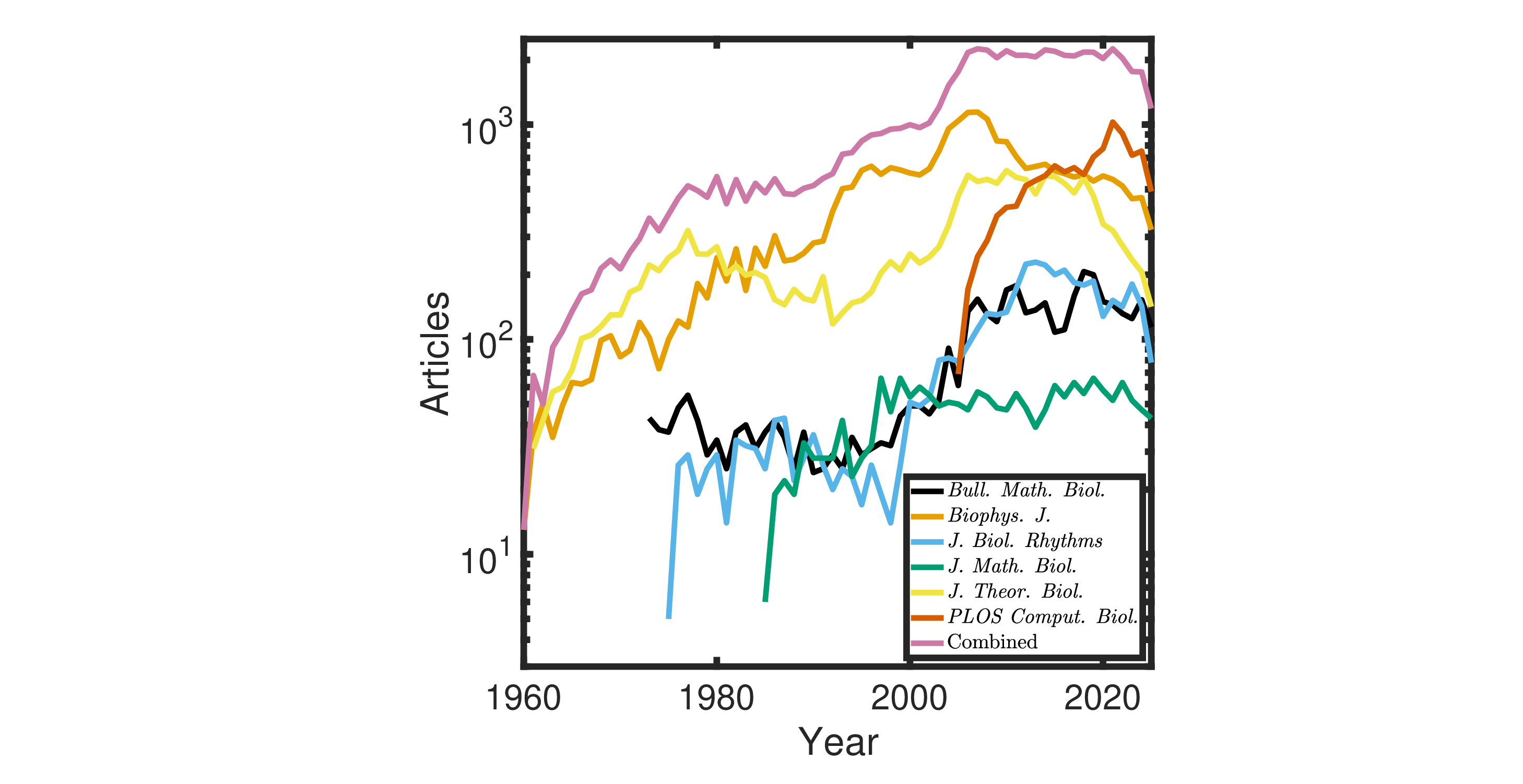}
    \caption{A growing number of mathematical models have been developed to describe biological systems and published in the literature. Journal articles over time published in mathematical biology-related journals, including the \textit{Bulletin of Mathematical Biology} (black), \textit{Biophysical Journal} (orange), \textit{Journal of Biological Rhythms} (blue), \textit{Journal of Mathematical Biology} (green), \textit{Journal of Theoretical Biology} (yellow), and \textit{PLOS Computational Biology} (orange) as indexed through results for each journal on PubMed. The combined total is shown in pink.}
    \label{fig:articles}
\end{figure}

 A huge number of mathematical models of biological systems are being developed and published each year (see Figure \ref{fig:articles}). To get a \textbf{crude and imperfect} estimate of the lower bound for how many articles are comparing across model structures, we went to each of the journals' websites that we considered and searched for ``Akaike Information Criterion,'' to give a lower bound (see Methods). Akaike Information Criterion is a classical and widely used statistical approach to model selection \cite{akaike1998information}. Across all time, we found between 40 and 133 results for this search in more classical mathematical biology journals \footnote{We considered \textit{Bulletin of Mathematical Biology}, the \textit{Biophysical Journal}, the \textit{Journal of Biological Rhythms}, the \textit{Journal of Mathematical Biology}, and the \textit{Journal of Theoretical Biology} to be more focused on classical mathematical biology.}, but 2,649 results (out of 11,461 total results, as indexed on PubMed)  in \textit{PLOS Computational Biology}, which has more of a focus on data-driven modeling. It seems reasonable to conclude that at least 23\% of studies in \textit{PLOS Computational Biology} contain some degree of model selection, but comprehensive model structure comparison is far from common (see Methods for further details). 
 
Some modeling studies of biological and ecological systems have attempted to carefully consider the influence of changes in model structure on modeling results, as we do here. A 2006 study fit a suite of seven differential equation models to Chronic Wasting Disease (CWD) outbreak data to examine different modes of transmission of Chronic Wasting Disease in deer populations \cite{miller2006dynamics}, though there is some question as to whether all the model structures considered were structurally identifiable \cite{fitzgerald2025effect}. A similar study was performed by Lee and co-authors to examine the influence of five different model structures on epidemiological modeling and forecasting of Cholera transmission \cite{lee2017model}. A more recent study used Bayesian model selection to compare across 4,208 possible models of interactions between microbial communities \cite{karkaria2021automated}. A recent article used recurrent neural networks to avoid a combinatorial search over all possible model structures for gene regulatory network inference, but the method appears to require higher temporal sampling than the data set we will examine here \cite{shen2021finding}. Additionally, a large array of model selection methods have been developed for data-driven model discovery of partially observed systems and avoid a comparison of all possible model structures, but typically these methods are demonstrated on higher temporal sampling \cite{ribera2022model,stepaniants2024discovering,lu2022discovering,bakarji2023discovering}. Here, we take a \textit{maximalist} approach and grapple with a challenging regulatory network inference problem. We fit a large family of 13,824 Ordinary Differential Equation (ODE) models, each model corresponding to a distinct regulatory network structure, to an existing gene expression data set to infer the currently unknown structural elements of a regulatory network that underpins a developmental decision in the nematode \textit{Pristionchus pacificus}.  

%add this in. 

% Model selection is not yet common place across mathematical biology, and more efforts to examine sources of structural uncertainty are important. 

 %add in this ref and look at references 1-6 in this ref: https://www.nature.com/articles/s41467-021-23420-5

 %Note that applying model selection algorithms approaches would require estimating derivatives which would not work in this data set. 

The nematode \textit{Pristionchus pacificus} exhibits developmental plasticity of its mouth form and feeding behavior \cite{bento2010co}. Adults are characterized as either Eurystomatous or Stenostomatous based on the number of tooth-like denticles and the proportional width and depth of their mouths. The Eurystomatous morph can kill other nematodes for food or competitive advantage, whereas the Stenostomatous morph eat only bacteria \cite{lightfoot2021sex}. Environmental perturbations are known to influence the population-level phenotypic ratio \cite{werner2017environmental,werner2018adult,piskobulu2025high,lenuzzi2023influence,bento2010co,bose2012complex,ragsdale2013developmental}, and a decade of genetic analysis has revealed >30 genes that affect the mouth-form development \cite{sommer2017genetics}. Epistatic experiments have ordered some of these genes relative to each other and environmental inputs. However, the dynamics of gene regulation and regulatory connections between genes in the network remain poorly understood and are the source of the high degree of structural uncertainty in our modeling problem. 

%obligate bacterivore 

Two genes in the network have been identified as ``switch'' genes. When either gene is mutated, worms exhibit a 100\% penetrant mouth-form phenotype regardless of environmental conditions, and when these genes are overexpressed, worms exhibit the opposite phenotype \cite{ragsdale2013developmental,bui2018sulfotransferase,namdeo2018two}. These two genes code for enzymes: (\textit{eud-1}) encodes a sulfatase, whereas (\textit{seud-1}/\textit{sult-1}, hereafter referred to as \textit{sult-1}), encodes a sulfotransferase. Sulfotransferases and sulfatases are enzymes with opposite biochemical functions; Sulfotransferases add and sulfatases remove sulfate groups from molecular substrates within the cell. Sulfation affects the activity of signaling molecules, is required for the metabolism and lysosomal degradation of many compounds, and generates diverse extracellular ligands \cite{strott2002sulfonation,igreja2022role}. \textit{eud-1} and \textit{sult-1} expression have opposite effects on mouth-form development, indicating that the relative abundance of these enzymes and their unidentified substrates determines the mouth-form phenotype. Furthermore, a recent RNA-seq experiment revealed that \textit{eud-1} and \textit{sult-1} are the only genes in the network whose expression is dependent on the environment during the critical window for mouth-form development \cite{reich2025developmental}. Thus, understanding the regulatory connections of these two genes is critical to understanding the logic of mouth-form plasticity. As an experimentally tractable system where the components of the gene regulatory network are known, determining their connectivity can inform plasticity writ large.

% The addition or removal of sulfate groups by sulfotransferases and sulfatases, respectively, is a ubiquitous biochemical reaction that regulates molecular and cellular signaling. 

% Sulfation affects the activity of signaling molecules, is required for the metabolism and lysosomal degradation of many compounds, and generates diverse extracellular ligands (cite Strott, Endocrine Reviews, 2002; Igreja & Sommer Frontiers Mol Biol 2022). 
Suppressor screens have identified several genes that operate downstream of \textit{eud-1} and \textit{sult-1}, although the structure and regulatory connections between them are also not well understood. The nuclear hormone receptor  \textit{nhr-40} is particularly interesting because it is a transcription factor, and it yields opposite phenotypes depending on the mutation (i.e., loss-of-function and gain-of-function alleles). To keep our analysis computationally tractable on a high-performance computational cluster, we explore a large family of effective regulatory network models involving \textit{eud-1}, \textit{sult-1}, and \textit{nhr-40}, which are thought to be the key players of the regulatory network underpinning the developmental decision in the nematode \textit{P. pacificus}. Beyond the developmental biology questions we pursue here, our general approach to mechanistic inference from typical experimental data can be applied to modeling contexts across biology where structural uncertainty is an issue and the properties of the experimental data makes algorithmic model selection infeasible.

% We then assess the model fits and analyze which model features are shared across those that best fit the data, in an attempt to resolve features of the gene regulatory network underpinning the mouth-form decision in \textit{Pristionchus pacificus} (Figure \ref{fig:1}). 

\begin{figure}[H]
    \centering
    \includegraphics[scale=0.2]{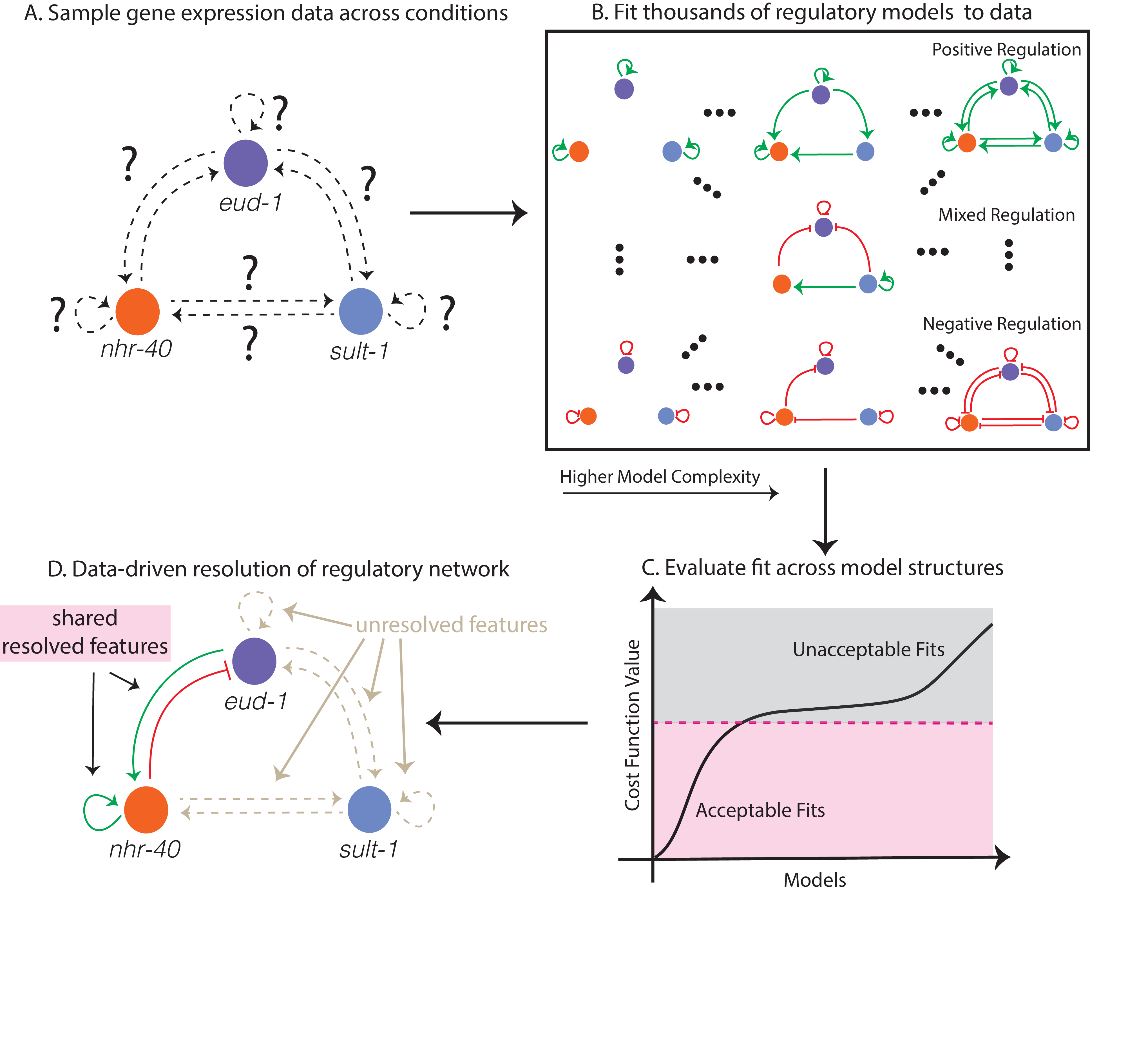}
    \caption{A: The regulatory network that underpins the dimorphic developmental mouthform decision in \textit{P. pacificus} is thought to involve \textit{eud-1}, \textit{nhr-40}, \textit{sult-1}, but the regulatory structure of the network is unknown. We use gene expression data sampled in three different experiments to identify key features of the regulatory network. B: We fit a family of ODE models to the gene expression data from the experiments. C: We assess the model fits to data across model structures. D: We then find the \textit{model set}, or collections of models with shared regulatory network features that best fit the data.}
    \label{fig:2}
\end{figure}

% As input we used RNA-seq data from three experimental conditions sampled across four time points in development: wild type grown on agar (95-100 percent Eu), \textit{eud-1} knock-out grown on agar (100 percent St), and \textit{sult-1} knock-out grown on agar (100 percent Eu).

We fit 13,824 distinct ODE model structures---each corresponding to a distinct effective regulatory network involving \textit{eud-1}, \textit{sult-1} and \textit{nhr-40}, to RNA-seq data from three experimental conditions (see Figure \ref{fig:2} and Sections \ref{data}, \ref{models}, \ref{structural}, \ref{estimation}). Using our maximalist computational approach, we resolve some features of the effective regulatory network, but we cannot completely distinguish between models given the data. We identify \textit{model sets}, or models with shared regulatory features that generate acceptable fits to each dataset (see Figure \ref{fig:2} and Section \ref{sets}). Similarly, we can rule out a host of possible regulatory network structures (see Figure \ref{fig:decision_tree}). We find a highly interconnected regulatory network structure dominated by positive regulation that can explain the data across each of the three experimental conditions, suggesting a \textit{possible} mechanism that underpins the mouth-form decision in \textit{Pristionchus pacificus}. More work is needed to experimentally verify the proposed mechanism and, more broadly, refine our model sets. The results we find here could be used to guide experiments aimed at achieving a mechanistic understanding of regulation during developmental decisions. This study examines \textit{practical indistinguishability}, a generalized form of sensitivity analysis across model structure given a data set, which is rarely considered in modeling studies but is increasingly important given the growing number of models being created. Our analysis allows us to probe the limits of what we can mechanistically infer, and is broadly applicable across scientific disciplines, including biology, chemistry, and physics. 

%% REMOVED rednance This analysis allows us to understand the sensitivity of our results as a function of model structure and the ability to differentiate between model structures as a function of the existing biological data. 

\section{Results}
\label{results}
Here, we begin our systematic search across 13,824 model structures using three experimental data sets. The data sets include the normalized RNA-seq reads for \textit{eud-1}, \textit{sult-1}, and \textit{nhr-40} sampled at six time points across development in three experimental conditions: wild type grown on agar, \textit{eud-1} knock-out grown on agar, and \textit{sult-1} knock-out grown on agar \cite{reich2025developmental} (see Sections \ref{data} and \ref{models}). We refer to these experiments as the wild type experiment, the \textit{eud-1} KO experiment, and the \textit{sult-1} KO experiment, respectively. As only the mRNA states were experimentally measured, we do not have any data for the associated protein states, and therefore, our model contains \textit{hidden states}. Systems with hidden states are known as \textit{partially-observed} and can exhibit \textit{structural identifiability issues}, meaning multiple parameter values in a model result in exactly the same dynamics of the measured states. In the context of parameter estimation, structural identifiability issues can result in non-unique fits to the data. 

We tested representatives from our large family of models for structural identifiability using STRIKE-GOLDD, and found they were not structurally identifiable, due to the presence of a scaling symmetry \cite{massonis2020finding}. Scaling symmetries are a type of algebraic transformation of parameters and variables of a model that leave the model invariant, meaning it produces the same dynamic behavior in the observables. Consider the following example model from our family of regulatory network models with the scaling symmetry highlighted in blue. From visual inspection, the scaling symmetry cancels from each equation of the model, meaning any value of $\xi$ produces the same mRNA dynamics. 

% An individual model is structurally identifiable if unique parameter values produce the observed output, and this is a data-independent algebraic property of the model structure. 

    \begin{align}
   \text{[eud-1 mRNA], Obs.:} & \dfrac{dM_e}{dt} = \alpha_1 + k_{t_{e_1}}\left( \dfrac{\textcolor{blue}{\xi}N}{\textcolor{blue}{\xi}K_{1} + \textcolor{blue}{\xi}N} \right)\left( \dfrac{\textcolor{blue}{\xi}S}{\textcolor{blue}{\xi}K_{4} + \textcolor{blue}{\xi}S} \right) -k_{-m} M_e, &M_e(0) = M_{eo} \label{eqni1} \\
    \text{[EUD-1 protein], Unobs.:} &\textcolor{blue}{\xi}\dfrac{dE}{dt} =  \textcolor{blue}{\xi}k_t M_e - \textcolor{blue}{\xi}k_{-p}\hat{E}, &E(0) = E_o \\
    \text{[nhr-40 mRNA], Obs.:} & \dfrac{dM_n}{dt} = \alpha_2 + k_{t_{n_1}}\left( \dfrac{\textcolor{blue}{\xi}N}{\textcolor{blue}{\xi}K_2 + \textcolor{blue}{\xi}N} \right) -k_{-m} M_n,&M_n(0) = M_{no} \\ 
    \text{[NHR-40 protein], Unobs:} &\textcolor{blue}{\xi}\dfrac{dN}{dt} =  \textcolor{blue}{\xi}k_t M_n - \textcolor{blue}{\xi} k_{-p}N,&N(0) = N_o \\
    \text{[sult-1 mRNA], Obs.:} &\dfrac{dM_s}{dt} = \alpha_3 + k_{t_{s_1}} \left( \dfrac{\textcolor{blue}{\xi}E}{\textcolor{blue}{\xi}K_{3} + \textcolor{blue}{\xi}E} \right)^2- k_{-m}M_s,&M_s(0) = M_{so} \\
    \text{[SULT-1 protein], Unobs.:}&\textcolor{blue}{\xi}\dfrac{dS}{dt} =  \textcolor{blue}{\xi}k_t M_s - \textcolor{blue}{\xi}k_{-p}S,&S(0) = S_o, \label{eqni2} 
\end{align}

%     \begin{align*}
%    \{Obs.\} ~~ & \dfrac{dM_e}{dt} = k_{t_{e_1}}\left( \dfrac{\color{blue}{e^{\epsilon}}\color{black}N}{\color{blue}{e^{\epsilon}}\color{black}K_{1} + \color{blue}{e^{\epsilon}}\color{black}N} \right) -k_{-m} M_e, ~~~~~~~~~~M_e(0) = M_{eo} \\
%     &\color{blue}{e^{\epsilon}}\color{black}\dfrac{dE}{dt} =  \color{blue}{e^{\epsilon}}\color{black} k_t M_e - \color{blue}{e^{\epsilon}}\color{black} k_{-p}E,~~~~~~~~~~~~~~~~~~~~~~~~~~ E(0) = E_o \\
%     \{Obs.\} ~~ & \dfrac{dM_n}{dt} = k_{t_{n_1}}\left( \dfrac{\color{blue}{e^{\epsilon}}\color{black}E}{\color{blue}{e^{\epsilon}}\color{black}K_2 + \color{blue}{e^{\epsilon}}\color{black}E} \right) -k_{-m} M_n,~~~~~~~~~~M_n(0) = M_{no} \\ 
%     &\color{blue}{e^{\epsilon}}\color{black}\dfrac{dN}{dt} =  \color{blue}{e^{\epsilon}}\color{black}k_t M_n - \color{blue}{e^{\epsilon}}\color{black}k_{-p}N,~~~~~~~~~~~~~~~~~~~~~~~~~~ N(0) = N_o \\
%     \{Obs.\} ~~ & \dfrac{dM_s}{dt} = k_{t_{s_1}} \left( \dfrac{\color{blue}{e^{\epsilon}}\color{black}K_{3}}{\color{blue}{e^{\epsilon}}\color{black}K_{3} + \color{blue}{e^{\epsilon}}\color{black}S} \right)- k_{-m}M_s,~~~~~~~~~~~ M_s(0) = M_{so} \\
%     &\color{blue}{e^{\epsilon}}\color{black}\dfrac{dS}{dt} =  \color{blue}{e^{\epsilon}}\color{black}k_t M_s - \color{blue}{e^{\epsilon}}\color{black}k_{-p}S,~~~~~~~~~~~~~~~~~~~~~~~~~~~ S(0) = S_o. 
% \end{align*}

If we attempted to parameterize the model without addressing the scaling symmetry, the estimated parameters would not be unique. Additionally, this identifiability issue occurs in all the models considered, as the difference between the 13,824 models is the form of the mRNA production term. For example, if we consider a model with a similar structure to Eqs. (\ref{eqni1})-(\ref{eqni2}) that includes additional positive regulation of \textit{nhr-40} by EUD-1, the same scaling symmetry impacts the dynamics of \textit{nhr-40}. 

    \begin{align*}
   % \text{[eud-1 mRNA], Obs.:} & \dfrac{dM_e}{dt} = \alpha_1 + k_{t_{e_1}}\left( \dfrac{\textcolor{blue}{\xi}N}{\textcolor{blue}{\xi}K_{1} + \textcolor{blue}{\xi}N} \right)\left( \dfrac{\textcolor{blue}{\xi}S}{\textcolor{blue}{\xi}K_{4} + \textcolor{blue}{\xi}S} \right) -k_{-m} M_e, &M_e(0) = M_{eo} \label{eq1} \\
   %  \text{[EUD-1 protein], Unobs.:} &\textcolor{blue}{\xi}\dfrac{dE}{dt} =  \textcolor{blue}{\xi}k_t M_e - \textcolor{blue}{\xi}k_{-p}\hat{E}, &\hat{E}(0) = \hat{E_o} \\
    \text{[nhr-40 mRNA], Obs.:} & \dfrac{dM_n}{dt} = \alpha_2 + k_{t_{n_1}}\left( \dfrac{\textcolor{blue}{\xi}N}{\textcolor{blue}{\xi}K_2 + \textcolor{blue}{\xi}N} \right)\left( \dfrac{\textcolor{blue}{\xi}E}{\textcolor{blue}{\xi}K_5 + \textcolor{blue}{\xi}E} \right) -k_{-m} M_n,&M_n(0) = M_{no} \\ 
    \text{[NHR-40 protein], Unobs:} &\textcolor{blue}{\xi}\dfrac{dN}{dt} =  \textcolor{blue}{\xi}k_t M_n - \textcolor{blue}{\xi} k_{-p}N,&N(0) = N_o 
    % \text{[sult-1 mRNA], Obs.:} &\dfrac{dM_s}{dt} = \alpha_3 + k_{t_{s_1}} \left( \dfrac{\textcolor{blue}{\xi}E}{\textcolor{blue}{\xi}K_{3} + \textcolor{blue}{\xi}E} \right)^2- k_{-m}M_s,&M_s(0) = M_{so} \\
    % \text{[SULT-1 protein], Unobs.:}&\textcolor{blue}{\xi}\dfrac{dS}{dt} =  \textcolor{blue}{\xi}k_t M_s - \textcolor{blue}{\xi}k_{-p}S,&S(0) = S_o, \label{eq2} 
\end{align*}

To eliminate the structural identifiability issue and break the scaling symmetry, we set $\xi=\dfrac{1}{k_t}$, which rescales the protein states and the half-max parameters by the translation rate. This is reasonable, as we know that the translation rate must be non-zero and has the benefit of reducing the dimension of the parameter space by one. We then checked a subset of the reparameterized models to ensure the resulting models were structurally identifiable using the package \texttt{StructuralIdentifiability.jl} \cite{structidjl}. For example, Eqs. (\ref{eqsi1})-(\ref{eqsi2}) are a structurally identifiable version of Eqs. (\ref{eqni1})-(\ref{eqni2}). We corrected all 13,824 models for this scaling symmetry ahead of parameter estimation (see Sections \ref{models} and \ref{structural}). Although we cannot verify the structural identifiability of all models due to computational limitations, removing at least this ubiquitous symmetry should improve the reliability of parameter estimation. 

%Identifiability analysis also encompasses \textit{practical identifiability} analysis, which examines if model parameters can be inferred uniquely \textit{given the data.} 

Before fitting our family of models to the experimental data, we performed a series of synthetic tests using Eqs. (\ref{eqsi1})-(\ref{eqsi2}) to understand the limits of inference in different data regimes. As Eqs. (\ref{eqsi1})-(\ref{eqsi2}) are structurally identifiable, any parameter identifiability issues should be practical in nature, meaning they emerge from the data quality rather than from symmetries in the structure of the equations.  To get a sense of how well we can recover the ground truth parameters, we sampled synthetic data from known parameters and varied the sampling frequency and noise level. We then refit the true model to the synthetic data sampled in different sampling and noise regimes. We also fit several ``misspecified model structures,'' or models with structural differences from the true model structure, to the synthetic data to understand the \textit{practical indistinguishability} of Model 11574 from other models across different data regimes. We refer to practical distinguishability as the ability to uniquely infer the structure given the data. ``Practically indistinguishable" was less examined in our search of the literature, perhaps due to a combination of its computational nature or the challenge of interpreting the results across multiple, competing model structures that emerge from structural uncertainty, though we did find some examples, such as \cite{karkaria2021automated}.

%across tens of thousands of model structures. 

\begin{figure}[H]
    \centering
    \includegraphics[scale=0.12]{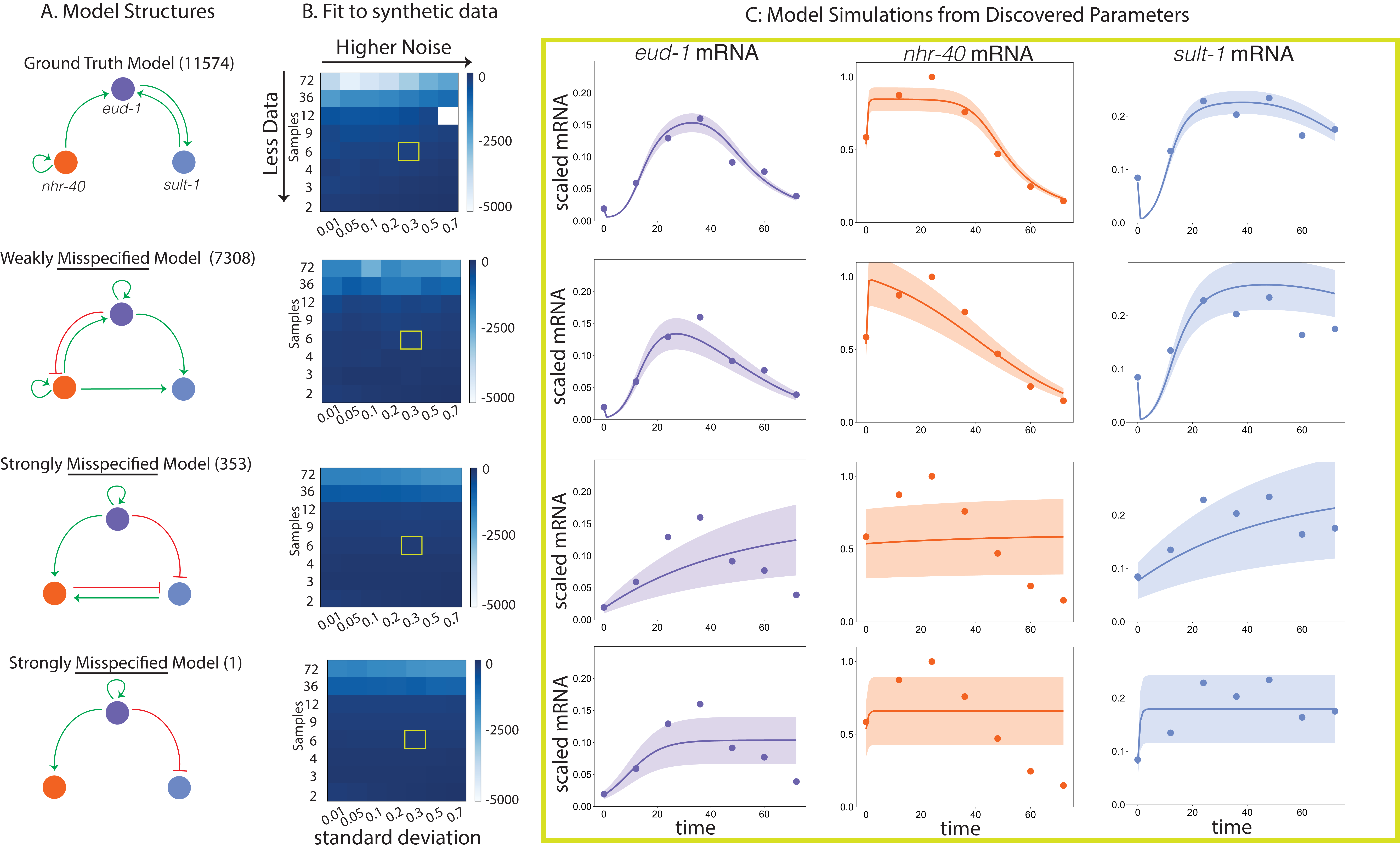}
    \caption{We generated synthetic data from the regulatory network Model 11574 (A, top) for a range of sampling rates and levels of noise, and then fit the true model structure along with several misspecified models to each of the synthetic data sets to understand recovery of the ground truth parameters and how misspecification in model structure influences the fits. Model 7308 (see A) shares many of the features of Model 11574, whereas Models 353 and 1 (see A) do not share any features with Model 11574. In the high sampling rate, low noise regime, the models are distinguishable as the true model has the lowest cost function value (top left corner of heat plots in B). However, in the approximate experimental data regime (denoted by yellow squares in B), the models are harder to distinguish from one another. Given the limited data and the relatively high noise, we do not expect to have practically identifiable parameters; however, certain model structures yield better fits to the data than others. A: Diagrams of the regulatory network structures for Models 11574, 7308, 353, and 1. B: Heat maps of the fits of Models 11574, 7308, 353, and 1 to synthetic data sampled from Model 11574 across a range of sampling rates and noise levels. The heat map is colored by the cost function value, lighter values are better fits to the data and darker values are poorer fits to the data. C: Forward simulations of the mRNA states of the models at the discovered parameter values estimated from the approximate experimental data regime (yellow box in B). The dynamics of \textit{eud-1} mRNA, \textit{nhr-40} mRNA, and \textit{sult-1} mRNA are shown in the purple, orange, and blue curves respectively. The corresponding synthetic data is shown in dots of the same color. The uncertainty (one standard deviation) is shown as the band of the corresponding color. Notice that Model 11574 and Model 7308 give a reasonable and similar fit to the synthetic data with experimentally-representative noise, whereas Models 353 and 1 give a very poor fit to the synthetic data sampled from Model 11574.}
    \label{fig:3}
\end{figure}

To determine how well we can recover ground truth parameters, we simulate synthetic data across a range of sampling rates corresponding to a few observations across the course of development to hourly sampling and a broad range of noise levels from Model 11574 (of 13,824) and attempt to fit the true model structure to the data (Figure \ref{fig:3}, B top). We select this model because it generates a good fit to the observed data, which we discuss in subsequent sections. Model 11574 has the form

    \begin{align}
   \text{[eud-1 mRNA], Obs.:} & \dfrac{dM_e}{dt} = \alpha_1+ k_{t_{e_1}}\left( \dfrac{\hat{N}}{\hat{K_{1}} + \hat{N}} \right)\left( \dfrac{\hat{S}}{\hat{K_{4}} + \hat{S}} \right) -k_{-m} M_e, &M_e(0) = M_{eo} \label{eqsi1} \\
    \text{[scaled EUD-1 protein], Unobs.:} &\dfrac{d\hat{E}}{dt} =  M_e - k_{-p}\hat{E}, &\hat{E}(0) = \hat{E_o} \\
    \text{[nhr-40 mRNA], Obs.:} & \dfrac{dM_n}{dt} = \alpha_2+k_{t_{n_1}}\left( \dfrac{\hat{N}}{\hat{K_2} + \hat{N}} \right) -k_{-m} M_n,&M_n(0) = M_{no} \\ 
    \text{[scaled NHR-40 protein], Unobs.:} &\dfrac{d\hat{N}}{dt} =  M_n - k_{-p}\hat{N},&\hat{N}(0) = \hat{N_o} \\
    \text{[sult-1 mRNA], Obs.:} &\dfrac{dM_s}{dt} = \alpha_3+k_{t_{s_1}} \left( \dfrac{\hat{E}}{\hat{K_{3}} + \hat{E}} \right)^2- k_{-m}M_s,&M_s(0) = M_{so} \\
    \text{[scaled SULT-1 protein], Unobs.:}&\dfrac{d\hat{S}}{dt} =  M_s - k_{-p}\hat{S},&\hat{S}(0) = \hat{S_o}, \label{eqsi2} 
\end{align}

where $M_e$, $\hat{E}$, $M_n$, $\hat{N}$, $M_s$, and $\hat{S}$ are the concentration of \textit{eud-1} mRNA, the scaled concentration of EUD-1 protein, the concentration of \textit{nhr-40} mRNA, the scaled concentration of NHR-40 protein, the concentration of \textit{sult-1} mRNA, and the scaled concentration of SULT-1 protein, respectively. As described above, the re-scaling of the variables $E$, $N$, and $S$ and the half-max parameters by the respective translation rate removes a scaling symmetry and ensures the parameters in this model are structurally identifiable. 

To analyze the impact of experimental noise on the practical indistinguishability of this model from others, we needed to choose an error model. We observed that the noise was higher when the mean of the data was higher, potentially indicating the presence of signal-dependent noise, which is commonly observed in biology, for example, see  \cite{furusawa2005ubiquity}. Based on this, a multiplicative error model is the simplest reasonable model we can assume, as there was not enough data to empirically estimate the error distribution and there was no straightforward first-principles approach based on knowledge of the system. We assumed a data-centered Gaussian with standard deviation proportional to the mean and generated synthetic data for a wide range of sampling rates (row in Figure \ref{fig:3}B) and noise levels (column in Figure \ref{fig:3}B). More negative log-likelihood values (lighter colors) correspond to better fits (see Methods). Using this error model and the known sampling rate, we can approximate the experimental data regime (yellow box, Figure \ref{fig:3}). When we fit the known model structure to synthetic data in the approximate experimental data regime, we get a reasonable fit to the data, though the exact parameter values are not recovered (Figure \ref{fig:3}, B top, yellow box and C top). Based on these synthetic tests, we do not expect each individual model to have practically identifiable parameters. 

To understand if we can uniquely infer the structure of a given model from data sampled in the relevant experimental regime---or how practically distinguishable Model 11574 is from other model structures, we explore fitting misspecified models to the synthetic data. We select three alternative structures or misspecified models to fit to the synthetic data generated by Model 11574. We selected Models 7308, 353, and 1 of the 13,824 structures, as they vary in how closely related their structure is to Model 11574, the true model structure in this synthetic test (Figure \ref{fig:3} A). Model 7308 and Model 11574 share several regulatory features, whereas Models 353 and 1 have completely different regulatory structures from Model 11574 and vary in their complexity (see Figure \ref{fig:3}A). Model 353 is much more complex than Model 1. In the relevant experimental data regime, we see that Model 7308 gives a reasonable fit to synthetic data generated from Model 11574 and has relatively tightly constrained uncertainty (Figure \ref{fig:3} C, second row), whereas Models 353 and 1 give a much poorer and more uncertain fit (Figure \ref{fig:3} C, third and fourth rows). In particular, the strongly mispecified models poorly describe the dynamics of \textit{nhr-40} mRNA over the entire course of development, as the uncertainty covers the entire possible range of the data. While we do not expect the parameters to be practically identifiable in the approximate experimental data regime, it is clear that some model structures are capable of generating a reasonable fit to the data, while others are not. Given the properties of the experimental data, we expect to be able to resolve some features of the network structure partially, but we do not expect this system to be practically distinguishable from all considered models.

\begin{figure}[H]
    
    \begin{center}
        
    \includegraphics[scale=0.18]{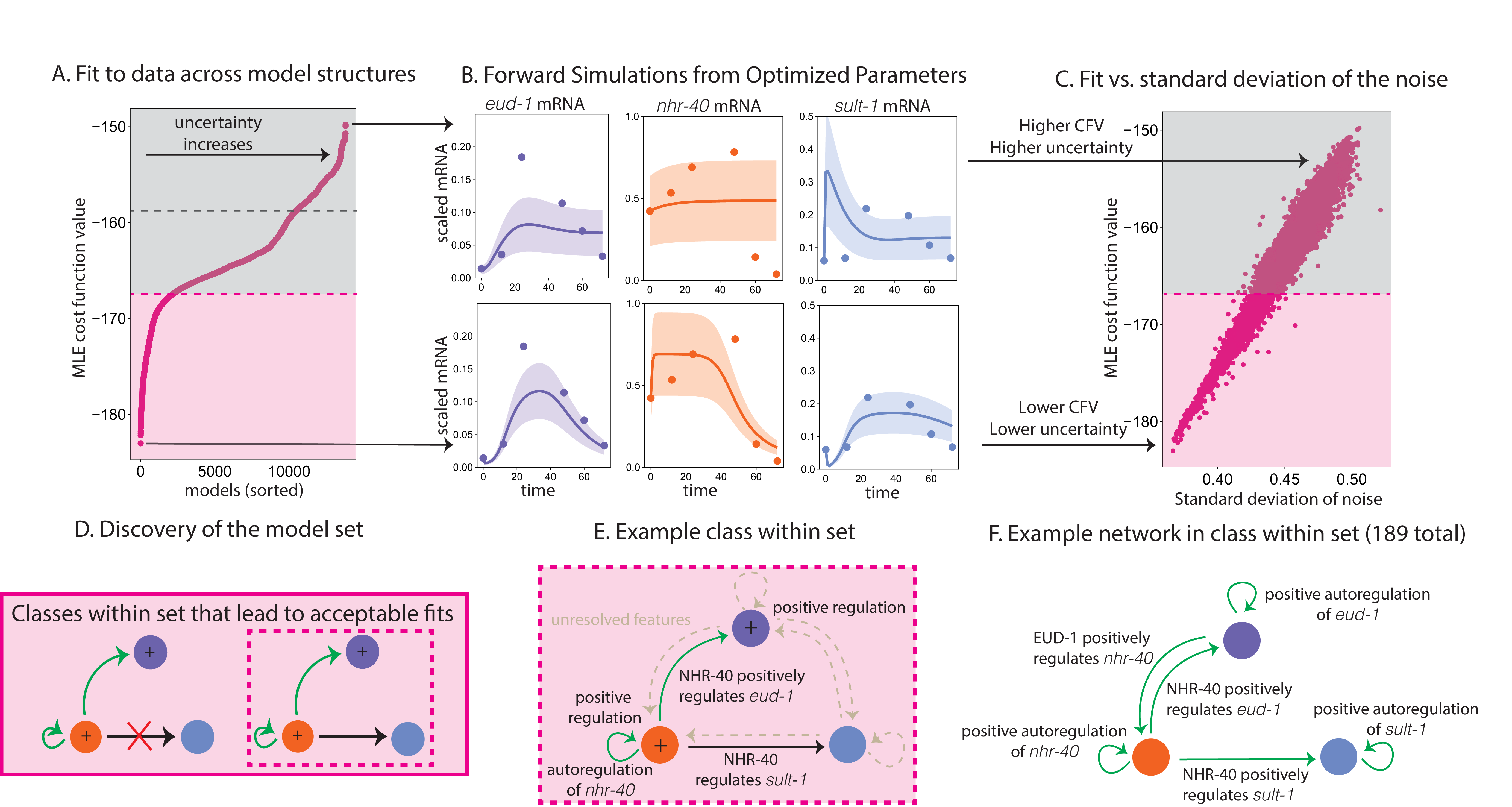}
    \end{center}
    \caption{A: We fit 13,824 models corresponding to different gene regulatory network structures to WT \textit{eud-1}, \textit{nhr-40}, \textit{sult-1} gene expression. Fitting each mathematical model to the data results in a cost function value. We sort the models by their cost function value. The curve that results has a noticeable plateau after about 1000 models (pink).  B: Forward simulations from the discovered parameters for a well-fit (bottom) and a poor-fitting model (top). The dynamics of \textit{eud-1} mRNA (purple), \textit{nhr-40} mRNA (orange), \textit{sult-1} mRNA (blue). The uncertainty (one standard deviation) is shown as the band of the corresponding color. C: Estimated standard deviation of the noise vs. the cost function value. As expected, poorer fits have a higher estimated noise. D: We use a decision tree classify acceptable fits (pink in A) and unacceptable fits (gray in A) based on structural features of the 13,824 models. We find two classes of model structures that predict acceptable fits. The first class includes positive regulation of \textit{eud-1} and \textit{nhr-40}, regulation of \textit{eud-1} by NHR-40, autoregulation of \textit{nhr-40} and no regulation of \textit{sult-1} by NHR-40. The second class (pink dashed box) includes positive regulation of \textit{eud-1} and \textit{nhr-40}, regulation of \textit{eud-1} by NHR-40, autoregulation of \textit{nhr-40} and regulation of \textit{sult-1} by NHR-40. E: Resolved features of the second class (dashed pink square) of the model set (D) are shown in black and the unresolved features as shown as dashed brown arrows. F: An specific example regulatory network structure from the example class shown in E. This network features a high degree of positive autoregulation and positive cross-regulation.}

    % D: Left, legend of symbols used. Middle: an example of a network class discovered in our analysis. This network class has 189 distinct regulatory networks within which each shares positive regulation of \textit{eud-1} and \textit{nhr-40}, regulation of \textit{eud-1} by NHR-40, autoregulation of \textit{nhr-40}, and regulation of  \textit{sult-1} by NHR-40. Right: one of the 189 networks that occupy the network class. This network is the unified regulatory network we describe in Figure \ref{fig:4}. E: Discovery of the model set. We used a decision tree to classify acceptable fits (pink in A) from unacceptable fits (gray in A). We discovered network features shared by models that generate acceptable fits (pink) and unacceptable fits (gray). We also find several combinations of model features that do not neatly predict acceptable or unacceptable fits. This is likely due to the rounded nature of the plateau in A. Note that only the features used in classification are shown in this diagram. }
    \label{fig:4}
\end{figure}

To infer structural features of the regulatory network that underpins the developmental decision in a wild-type genetic background, we fit 13,824 different model structures to RNA-seq data from the wild-type experiment. A subset of around 1000 models gives a reasonable fit to data (Figure \ref{fig:4} A). After about 1000 models, the cost function curve over model structures plateaus, before increasing dramatically (see Figure \ref{fig:4} A). The best-fitting models give a reasonable description of the data, in that the estimate closely describes the data and the uncertainty is more constrained than other model structures in the family we considered (Figure \ref{fig:4}B, bottom). The model structures that result in the highest cost function values do not describe the data well and are highly uncertain (Figure \ref{fig:4}B, top). In particular, the estimate for \textit{nhr-40} mRNA is almost entirely unconstrained. Qualitatively, the cost function curve seems to have three distinct regions: models that describe the gene expression data reasonably well (pink in Figure \ref{fig:4} A), models that describe the gene expression data more poorly but are not dominated by uncertainty (gray before the dashed line in Figure \ref{fig:4} A), and models that do not capture the data and the uncertainty is as large as the dynamic range of the data (gray after the dashed line in Figure \ref{fig:4} A). As expected, there is a positive correlation between the model's cost-function-value and the estimate's uncertainty (Figure \ref{fig:4}C). 

To learn which characteristics distinguish better fitting models, we used a decision tree to classify models that produced an acceptable fit (Figure \ref{fig:4} A, pink) from those that produced an unacceptable fit (Figure \ref{fig:4} A, gray), using a broad array of 21 distinct structural features of the models (see Section \ref{sets}). We find that the models that generate the best, more certain fits to data are enriched in several model features (Figure \ref{fig:4} D, E, F ). In particular, NHR-40 likely regulates \textit{eud-1} and \textit{nhr-40} and sometimes \textit{sult-1}. We also found that \textit{eud-1} and \textit{nhr-40} are likely positively regulated, and \textit{nhr-40} is likely autoregulated. We found that 325 regulated networks across two classes that all generate an acceptable fit to the data (Figure \ref{fig:4}D). 

Notably, not all models that generate an acceptable fit to the data have these characteristics. Roughly 35\% of the model structures that produce an acceptable fit share these features (see Figure \ref{fig:decision_tree}). The analysis also revealed a broad array of 21 model structures that predictably generate unacceptable fits to the data (see Figure \ref{fig:decision_tree}). For some models, the decision tree could not predict whether a model would yield an acceptable or unacceptable fit (see Figure \ref{fig:decision_tree}). This is expected because the plateau of the cost function value curve (Figure \ref{fig:3} A) is relatively rounded, so models on either side of the threshold we selected likely share common features. A handful of model structures resulted in many of the misclassifications (see Figure \ref{fig:decision_tree}). These structures are closely related to network structures that produce both acceptable and unacceptable fits to the data. 

We then repeated the model fitting procedure for \textit{eud-1} KO experiment and \textit{sult-1} KO experiment. To model the knock-out experiments, we removed all production terms that involved the protein associated with the gene that was knocked out, reducing the number of possible production terms (see Methods).

\begin{figure}[H]
    \centering
    \includegraphics[scale=0.16]{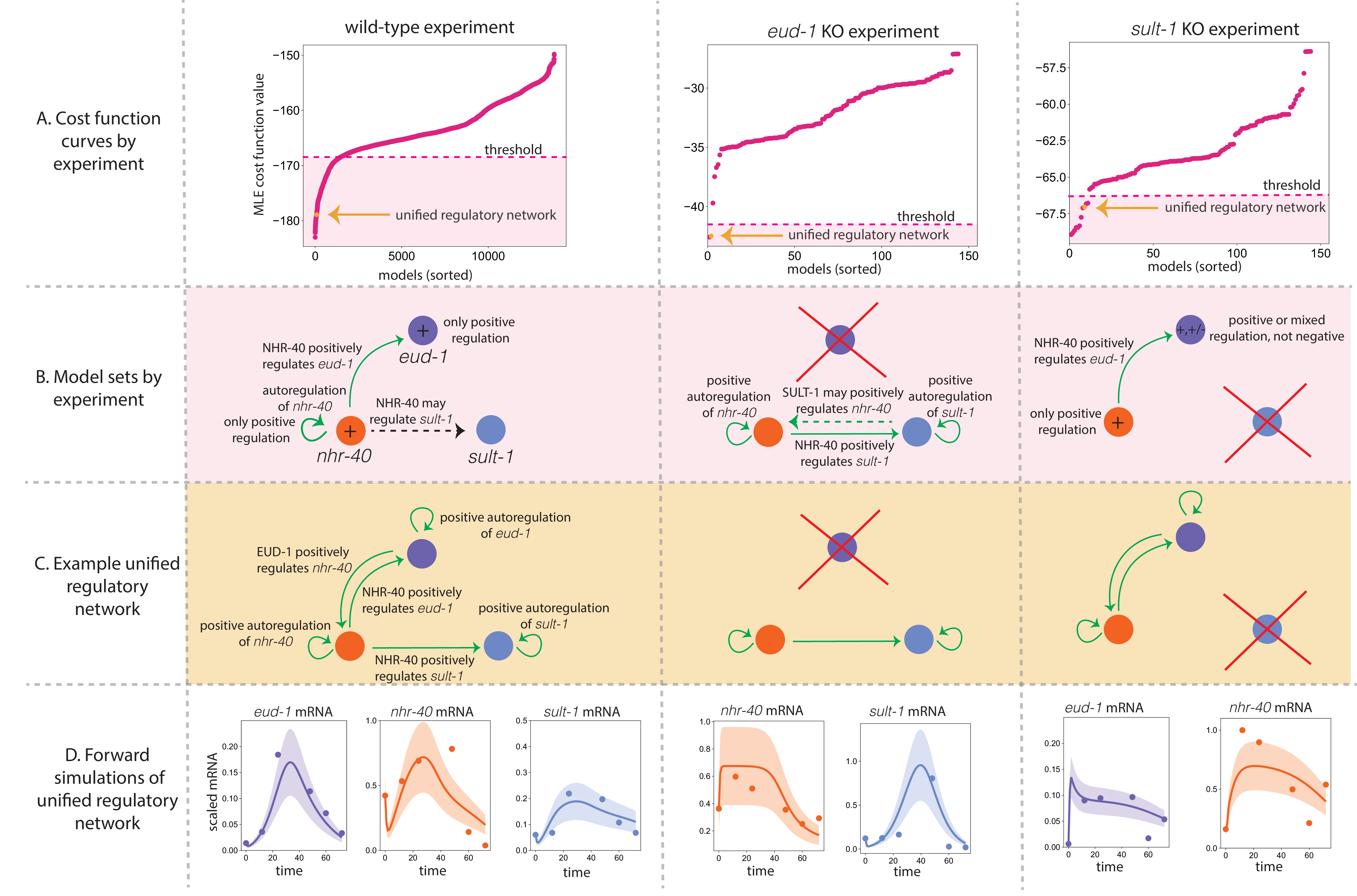}
    \caption{A: Cost function curves sorted by value for the WT experiment, the \textit{eud-1} knock-out experiment, and the \textit{sult-1} knock-out experiment. Pink shaded regions indicate the best fits up to a plateau in the curve. B: Commonality in the regulatory network structure for the best fitting models (pink shaded region in A). C: An example unified regulatory network. The model structure appears in (A) in orange. D: Model simulations of the unified regulatory mechanism across experiments at optimized parameter values (only one replicate shown). Scaled \textit{eud-1} mRNA, \textit{nhr-40} mRNA, \textit{sult-1} mRNA expression is shown as purple, orange, and blue dots, respectively. Model simulations at optimization parameter values are shown as a solid line in the corresponding color. The uncertainty (one standard deviations) is shown as the band of the corresponding color.}
    \label{fig:5}
\end{figure}

When we fit a wide range of model structures to the \textit{eud-1} KO experiment, we see that only a handful of models achieve a good fit (see Figure \ref{fig:5}A, middle), and the cost function curve exhibits a large gap to models with poorer fits. The two models with the lowest cost function values have very similar model structures. In particular, \textit{nhr-40} is positively autoregulated and \textit{sult-1} is positively regulated by NHR-40 and positively autoregulated. The difference between the models is whether or not SULT-1 positively regulates \textit{nhr-40}. When we fit a wide range of model structures to \textit{sult-1} KO experiment, we notice that about 10 models seem to give a reasonable fit to the data before the cost function curve plateaus (see Figure \ref{fig:5}A, right). These models tend to include positive regulation of \textit{eud-1} by NHR-40, autoregulation of \textit{eud-1} (positive or negative), and either positive autoregulation of \textit{nhr-40} or positive regulation of \textit{nhr-40} by EUD-1. We then probe the model sets associated with each of the experiments (Figure \ref{fig:5} A, pink) for a common regulatory network structure that can explain all three experiments. We find a model structure (Figure \ref{fig:5} A, orange dot and Figure \ref{fig:5} C) with strong commonalities to the model we did synthetic testing on. Both models feature a high degree of positive regulation of the network and include regulation of \textit{eud-1} by NHR-40 and autoregulation of \textit{nhr-40}. The unified model structure has the form 

    \begin{align}
   \text{[eud-1 mRNA], Obs.:} & \dfrac{dM_e}{dt} = \alpha_1 + k_{t_{e_1}}\left( \dfrac{\hat{E}}{\hat{K_{1}} + \hat{E}} \right)\left( \dfrac{\hat{N}}{\hat{K_{4}} + \hat{N}} \right) -k_{-m} M_e, &M_e(0) = M_{eo} \label{eq1} \\
    \text{[scaled EUD-1 protein], Unobs.:} &\dfrac{d\hat{E}}{dt} =  M_e - k_{-p}\hat{E}, &\hat{E}(0) = \hat{E_o} \label{eq2} \\
    \text{[nhr-40 mRNA], Obs.:} & \dfrac{dM_n}{dt} = \alpha_2 + k_{t_{n_1}}\left( \dfrac{\hat{E}}{\hat{K_2} + \hat{E}} \right)\left( \dfrac{\hat{N}}{\hat{K_{5}} + \hat{N}} \right) -k_{-m} M_n,&M_n(0) = M_{no} \label{eq3} \\ 
    \text{[scaled NHR-40 protein], Unobs.:} &\dfrac{d\hat{N}}{dt} =  M_n - k_{-p}\hat{N},&\hat{N}(0) = \hat{N_o} \label{eq4} \\
    \text{[sult-1 mRNA], Obs.:} &\dfrac{dM_s}{dt} = \alpha_3 + k_{t_{s_1}} \left( \dfrac{\hat{N}}{\hat{K_{3}} + \hat{N}} \right)\left( \dfrac{\hat{S}}{\hat{K_{6}} + \hat{S}} \right)- k_{-m}M_s,&M_s(0) = M_{so} \label{eq5} \\
    \text{[scaled SULT-1 protein], Unobs.:}&\dfrac{d\hat{S}}{dt} =  M_s - k_{-p}\hat{S},&\hat{S}(0) = \hat{S_o}, \label{eq6} 
\end{align}

where $M_e$, $\hat{E}$, $M_n$, $\hat{N}$, $M_s$, and $\hat{S}$ is the concentration of \textit{eud-1} mRNA, the scaled concentration of EUD-1 protein, the concentration of \textit{nhr-40} mRNA, the scaled concentration of NHR-40 protein, the concentration of \textit{sult-1} mRNA, and the scaled concentration of SULT-1 protein, respectively. As predicted by the feature importance analysis using the decision tree decision tree, the model is dominated by positive regulation and features regulation of \textit{eud-1} by NHR-40 and autoregulation of \textit{nhr-40}. When \textit{sult-1} is knocked out, the model reduces to Eqs. (\ref{eq1})-(\ref{eq4}). When \textit{eud-1} is knocked out, the model reduces to the following equations, 

    \begin{align}
   \text{[eud-1 mRNA], Obs.:} & \dfrac{dM_e}{dt} = \alpha_1+ k_{t_{e_1}}\left( \dfrac{\hat{E}}{\hat{K_{1}} + \hat{E}} \right)\left( \dfrac{\hat{N}}{\hat{K_{4}} + \hat{N}} \right) -k_{-m} M_e, &M_e(0) = M_{eo} \label{eq12} \\
    \text{[scaled EUD-1 protein], Unobs.:} &\dfrac{d\hat{E}}{dt} =  M_e - k_{-p}\hat{E}, &\hat{E}(0) = \hat{E_o} \label{eq22} \\
    \text{[nhr-40 mRNA], Obs.:} & \dfrac{dM_n}{dt} = \alpha_2 + k_{t_{n_1}}\left( \dfrac{\hat{N}}{\hat{K_{5}} + \hat{N}} \right) -k_{-m} M_n,&M_n(0) = M_{no} \label{eq32} \\ 
    \text{[scaled NHR-40 protein], Unobs.:} &\dfrac{d\hat{N}}{dt} =  M_n - k_{-p}\hat{N},&\hat{N}(0) = \hat{N_o} \label{eq42}.  
\end{align}

This implies that \textit{nhr-40} is autoregulated in the absence of \textit{eud-1}. This would explain the somewhat perplexing result that \textit{sult-1} expression is significantly increased in the \textit{eud-1} knockout - even though neither \textit{eud-1} nor \textit{sult-1} is a transcription factor. Forward simulations of the model structures fit to the corresponding data are shown in Figure \ref{fig:5} D. 

Here, we attempted to infer the structure of a regulatory network that controls the dimorphic mouth form decision in \textit{P. pacficus} using RNA-seq data from three experiments. We fit a total of 13,824 models to the wild type experiment, and 144 models to each of the knock-out experiments, and found a model set, or combinations of features that are shared across the models that generate acceptable fits to the data, for each experiment (see Figure \ref{fig:5}B). Collating these model sets, we put forth a unified regulatory network structure that exists in each of the model sets associated with the experiment (see Figure \ref{fig:5}). The regulatory network model is relatively complex and features a high degree of positive autoregulation and positive cross-regulation (see Figure \ref{fig:5}C). It is important to note that this is one possible regulatory motif that explains the data reasonably well across experiments, but a very high degree of uncertainty still exists given the size of the model sets associated with the wild type and \textit{sult-1} knock out experiment---roughly 1000 models and 10 models, respectively. More experimental work is needed to refine our model sets and verify or exclude the possible regulatory network structure that we identified in our analysis. 

\section{Discussion}
 
Here we performed a case study of practical indistinguishability across 13,824 distinct model structures to infer features of a regulatory network that underpins a developmental decision in the nematode \textit{P. pacificus}. We found model sets, or collections of models with shared regulatory network features that best fit the data, for each of the three experiments (see Figures \ref{fig:4} and \ref{fig:5}). We then found a regulatory network in the intersection of the three model sets and is capable of describing the data across each of the experimental conditions (see Figure \ref{fig:5}C and D). This regulatory network exhibits a high degree of positive regulation and interconnectivity between the key regulators, \textit{eud-1}, \textit{sult-1}, and \textit{nhr-40}. There are several natural next steps to be addressed on the biological and mathematical side. 

To further refine our understanding of the underlying mechanism of the developmental decision based on what we learned in this \textit{in silico} analysis, there are several next steps to be addressed in future work. Across model sets, we found that NHR-40 was heavily involved in the regulation of the effective networks we discovered. A deeper understanding of the regulatory network could be found through an \textit{nhr-40} knock-out experiment, to understand the dynamics of the system in the absence of this key regulator. Additionally, one could use the discovered model sets to perform optimal experiment design to suggest new experiments that limit cost and maximize our ability to distinguish the models associated with each experiment. Possible informative experiments include more frequently sampled gene expression data, measurements of protein states, or applying other perturbations to system states that have an indirect effect on the genes of interest.

Mathematically, this analysis is a natural complement to emerging theoretical results in structural indistinguishability \cite{bortner2024graph,gross2019linear}. The analysis we presented here examined practical indistinguishability across non-linear model structures using experimental data. Mathematically, it would be of interest to extend the structural indistinguishability results to nonlinear systems and understand the connections between practical identifiability, or the ability to find unique parameters given the associated data, and practical indistinguishability. Structural indistinguishability is a data-independent approach to understanding which models can be distinguished from one another using the input-output equation or graph theoretic approaches \cite{bortner2024graph,gross2019linear}. These theoretical results are largely limited to linear models and cannot currently be applied to non-linear systems, such as Eqs. (\ref{eq1})-(\ref{eq2}) and do not consider the impact of experimental data on indistinguishability. A natural next step is to attempt to extend the theoretical results for linear systems to systems with non-linear forms, such as Eqs. (\ref{eq1})-(\ref{eq2}). Practical identifiability examines if unique parameters can be found \textit{given} the associated data. An additional future direction would be to extend frameworks for practical identifiability to examine and identify sources of structural uncertainty in the model, or identify additional terms that would retain practical identifiability given the data. 

% In the future, we hope to extend structural indistinguishability results to nonlinear systems and study interactions between structural and practical indistinguishability more in-depth. 

This analysis has several important biological and computational limitations. On the biological side, we assumed that the proteins EUD-1, NHR-40, and SULT-1 can directly or indirectly influence the production of \textit{eud-1}, \textit{nhr-40}, and \textit{sult-1} mRNA. We model the mRNA production terms as Hill Functions involving the proteins, as the intermediate regulatory steps are unknown, and biologically, little is known about the true transcriptional regulators of \textit{eud-1}, \textit{nhr-40}, and \textit{sult-1}. It is also important to note that the true regulatory network that controls the dimorphic mouth-form decision is more complex and expansive than the three genes we opted to model \cite{ragsdale2013developmental,bui2018sulfotransferase,namdeo2018two, sieriebriennikov2018developmental,serobyan2016chromatin,kieninger2016nuclear,sieriebriennikov2020conserved,moreno2019cilia,casasa2023mediator}. Therefore, the network regulatory features we uncovered may be an indirect rather than a direct effect. Nevertheless, our models represent a minimal and effective description of the regulatory network that underpins the dimorphic mouth-form decision in \textit{P. pacificus}. Part of the decision to limit our modeling scope to these three genes is based on the biological knowledge of their importance, but also to keep our analysis computationally tractable, given its combinatorial nature and heavy computational footprint. 

We also assume that the broad dynamics of the regulatory network are deterministic and can be described by a deterministic model. However, it is known that gene expression can be stochastic \cite{raj2008nature}. Our efforts are a first pass at developing a model for the underlying regulatory network, and inference of a stochastic system would not be possible without more sample time points and likely single-cell measurements. Given the highly conserved nature of this developmental pathway and the fact that we are analyzing bulk RNA-seq rather than data from single cells, it is perhaps reasonable to expect a relatively deterministic genetic signal at the organismal level.  

We also needed to make choices during optimization that may have some influence on the results. As we did not have good estimates for any of our parameters, we set the lower and upper bounds for our unscaled kinetic parameters to be 1x10$^{-5}$ and 1x10$^4$+1, respectively. Scaled model parameters, such as $$\hat{N}(0) = \dfrac{N(0)}{k_t},$$ were estimated between 1x10$^{-9}$ and 1x10$^8$+1. Narrower parameter spaces tended to lead to poor fits or fits in which parameters often hit the bounds during optimization. There is a trade-off between the size of the parameter space and our ability to efficiently search it using multi-start optimization (see Section \ref{estimation}). The fact that models with similar structures provide the best fits to the data provides some reassurance that we can navigate this space in a reasonable manner. In our analysis, we noticed that many of the best fits across model structures included a relatively high estimate for $\hat{N}(0)$, the initial condition for the scaled NHR-40 protein, on the order of magnitude of 1x10$^7$. The only way that $\hat{N}(0)$ could be on the order of magnitude of 1x10$^7$ is if $\hat{N}(0)$ is relatively large and the translation rate, $k_t$, is relatively small. It is feasible and perhaps even likely that additional biological processes not captured in our model are lowering the effective translation rate, such as a post-transcriptional modification or RNA degradation that is beyond the scope of the model and current understanding of the system. 

In the model sets associated with each of the experiments, NHR-40 was implicated as a regulator of the network, but this may be an artifact of key modeling assumptions we were forced to make due to the constraints in the data. Based on the experimental data, \textit{nhr-40} gene expression was much higher than either \textit{eud-1} or \textit{sult-1} in all experiments. In an attempt to balance the biological reality and limit the number of parameters we attempt to estimate, we did not attempt to estimate the initial condition for EUD-1 or SULT-1. We assume that if gene expression is relatively low, as it is for \textit{eud-1} or \textit{sult-1} across all experiments, then the associated protein concentration should also be relatively low. Therefore, we fixed the EUD-1 and SULT-1 protein initial condition to a low pre-set value during parameter estimation (see Methods). Our structural identifiability analysis suggested that the initial condition for the scaled NHR-40 protein is identifiable if it appeared in at least one of the mRNA production terms. For model structures that included NHR-40 as a regulator of the network, we sought to estimate the initial condition, $\hat{N}(0)$. It may not be surprising that models that best fit the data often included NHR-40 somewhere in the regulatory network, as these models included one additional parameter, $\hat{N}(0)$. However, we inferred more features of the regulatory network than just the presence of NHR-40 as a regulator in our analysis (see Figures \ref{fig:4}-\ref{fig:5}). 
 
 This work was relatively computationally heavy. After trial and error, we ran parameter estimation using 1000 seeds for each model. It is possible that the results could slightly change if we were able to re-run our analysis using 10,000, 100,000, or more seeds, though we obtained similar results when we spot-checked a handful of models using 10,000 seeds. Increasing to 100,000, or more seeds for all models was infeasible given our current computing resources and the number of model structures we wanted to probe. 

 An additional caveat relates to our choice of error model. The biological data are relatively noisy across replicates, especially when gene expression is higher. The number of replicates and time points - while standard, if not better than many biological data sets - did not enable us to determine the error model for the system in a data-driven manner, and there was no clear first principles approach to determining the error model using knowledge of the system. The data may also include some error in time or the ``x-direction,'' which can occur if a replicate is slightly ahead or behind others, and further complicates the error model. We did not attempt to correct for this possible temporal asynchrony. Ultimately, we assumed a multiplicative error model, largely based on qualitative features of the data, and optimized a maximum likelihood estimate (see Section \ref{estimation}), though we also found similar qualitative modeling results when we optimized a weighted least squares cost function and prescribed the weight to be inversely related to the variance across replicates. 

 Here, we attempted to probe the limits of mechanistic inference using data-driven modeling applied to biological data in a common yet challenging data regime. We found that we could partially resolve features of the effective regulatory network that controls the dimorphic mouth-form decision in \textit{P. pacficus}, but could not completely discern the structure of the regulatory network given the limitations of the data, our current understanding of the underlying biology, and our computational constraints. Our data-driven modeling results identify features of the regulatory network that generate acceptable fits to the known data and suggest future experiments, such as temporally sampling expression of \textit{eud-1} and \textit{sult-1} in a loss of function \textit{nhr-40} knockout over the course of development or using the discovered model sets to perform optimal experimental design to suggest sampling times, to further refine features of the regulatory network. In turn, our analysis can be iteratively run as new data becomes available, potentially narrowing the size of our model sets and improving our ability to distinguish model structures. Beyond the biological implications of our work, this analysis attempts to understand the sensitivity of our results as a function of model structure and probes what can be \textit{mechanistically} inferred based on typical biological data. These analyses can be applied to any system where structural uncertainty poses a challenge and the experimental data makes algorithmic model selection infeasible.

% Our modeling results give a new set of data-driven hypotheses concerning the structure of the regulatory network that can be experimentally probed. 

% In turn, our analysis can be iteratively run as new data becomes available, potentially narrowing the size of our model sets and improving our ability to distinguish model structures. Beyond the biological implications of our work, this analysis attempts to understand the sensitivity of our results as a function of model structure and probes what can be \textit{mechanistically} inferred based on typical biological data. These analyses can be applied to any system where structural uncertainty poses a challenge.

%DISCUSSION OF error model and errors in the x direction. 

%basic results of what we found by experiment

%paragraph describing possible regulatory network that is unified. 

%reasons for caution and nuance concerning the results. 
%-gene regulation is unclear, proteins are proxies. {DONE} 
%-this is a partial observation of a complex system, surely it may be more complex? {DONE}
%-N appeared in the model, perhaps not surprising but it did show up mutliple places. {DONE}
%-regulation type
%-limited by computational power and our ability to fully explore cost function landscape. 
%-parameter regime: translation rate must be very low, meaning something else must be involved. 
%-symmetries--only corrected for some of the symmetries. {DONE} 
%--y error vs. x error. 
%unknown error model. 
\section{Conclusion}

Here, we performed the most rigorous data-driven modeling possible based on data that is typical of biological experiments. We attempted to extract as much mechanism as possible from the data, while taking seriously the high degree of structural uncertainty inherent in the problem by fitting a very large family of ODE models to the data. We find that given the data, we are able to partially resolve features of the regulatory network, which provides a data-driven way to guide future experimentation and identify a unified regulatory network that can explain multiple experimental findings. More experimental work is needed to verify this regulatory network and, more broadly, zero in on the underlying mechanism that controls mouth-form plasticity, but this work helps to further our understanding of the system and probe the limits of inference using a maximalist computational approach.

\section{Methods \& Technical Notes}

Code associated with this project can be found at \url{https://github.com/cefitzg/mouthform_code}. Supercomputing code can be made available upon request. 

\subsection{Examination of model selection in mathematical biology journals}

To get a rough estimate of the number of articles that have been published in six mathematical biology journals, including the \textit{Bulletin of Mathematical Biology}, the \textit{Biophysical Journal}, the \textit{Journal of Biological Rhythms}, the \textit{Journal of Mathematical Biology}, the \textit{Journal of Theoretical Biology}, and \textit{PLOS Computational Biology}, we filtered PubMed results by journal and recorded the number of results across time. An example search can be found here, \url{https://pubmed.ncbi.nlm.nih.gov/?term=%22Journal+of+Theoretical+Biology%22%5BJournal%5D+}. 

To search for the term ``Akaike Information Criterion'' in each journal, we went to the website for each journal and searched for the term. An example search can be found here, \url{https://www.sciencedirect.com/search?qs=Akaike+Information+Criterion&pub=Journal+of+Theoretical+Biology&cid=272314}. We initially attempted to use PubMed to search for the term ``Akaike Information Criterion,'' but found many fewer results for each journal. 

\subsection{Data}

Here, we describe the experimental data in depth and the scaling we applied to the data for our analysis. 

\subsubsection{Data sets}
\label{data}
In our analysis, we used RNA-seq data from three experimental conditions: wild type \textit{Pristionchus pacificus} (reference line PS312), \textit{eud-1} knock-out (\textit{tu1069}), and \textit{sult-1} knock-out (\textit{tu1061}), all grown on NGM agar plates. Each condition included samples from six time points across development from embryo to adulthood (0 h, 12 h, 24 h, 48 h, 60 h, and 72 h post-bleach synchronization). The wild-type experiment included four biological replicates at each time point, whereas the two knock-out experiments used two replicates per time point. The data for this analysis were selected from a larger, recently published transcriptomic analysis in \textit{Pristionchus} \cite{reich2025developmental}.

%We used normalized gene expression results from DESeq2 \cite{love2014moderated} for model parameterization (see Section \ref{scaling}).

\subsubsection{Data scaling}
\label{scaling}
For each experiment, we scaled all gene expression data across time points and replicates by the maximum, which transforms the data between zero and one for each experiment. 

\subsection{Family of regulatory models descriptions}
\label{models}
To model the wild-type experiment, we used a family of models with the form 

    \begin{align}
   \text{[eud-1 mRNA]:} ~\dfrac{dM_e}{dt} &= \alpha_1 + k_{t_{e}}F_1(E,N,S,K_1,K_4)-k_{-m} M_e,\label{agar_model1}\\
    \text{[EUD-1 protein]:} ~\dfrac{dE}{dt} &= k_{t} M_e - k_{-p}E, \\
    \text{[nhr-40 mRNA]:} ~\dfrac{dM_n}{dt} &= \alpha_2+k_{t_{n}}F_2(E,N,S,K_2,K_5)-k_{-m} M_n, \\ 
    \text{[NHR-40 protein]:} ~\dfrac{dN}{dt} &= k_{t} M_n - k_{-p}N, \\
    \text{[sult-1 mRNA]:} ~\dfrac{dM_s}{dt} &= \alpha_3 + k_{t_{s}}F_3(E,N,S,K_3,K_6) -k_{-m} M_s, \\
    \text{[SULT-1 protein]:} ~\dfrac{dS}{dt} &= k_{t} M_s  - k_{-p}S. \label{agar_model2} 
\end{align}

where $F_i(E,N,S)$ for $i=1,2,3$ is defined as

\begin{align*}
    F_i(E,N,S) &=\\
    &\Biggl\{\left(\dfrac{E}{K_i+E}\right), \left(\dfrac{K_i}{K_i+E}\right),\left(\dfrac{N}{K_i+N}\right),\left(\dfrac{K_i}{K_i+N}\right),\left(\dfrac{S}{K_i+S}\right),\left(\dfrac{K_i}{K_i+S}\right),\\
    &\left(\dfrac{E}{K_i+E}\right)^2, \left(\dfrac{K_i}{K_i+E}\right)^2,\left(\dfrac{N}{K_i+N}\right)^2,\left(\dfrac{K_i}{K_i+N}\right)^2,\left(\dfrac{S}{K_i+S}\right)^2,\left(\dfrac{K_i}{K_i+S}\right)^2, \\
    &\left(\dfrac{E}{K_i+E}\right)\left(\dfrac{N}{K_{i+3}+N}\right), \left(\dfrac{E}{K_i+E}\right)\left(\dfrac{K_{i+3}}{K_{i+3}+N}\right),\left(\dfrac{E}{K_i+E}\right)\left(\dfrac{S}{K_{i+3}+S}\right), \\ &\left(\dfrac{E}{K_{i}+E}\right)\left(\dfrac{K_{i+3}}{K_{i+3}+S}\right),\left(\dfrac{K_i}{K_i+E}\right)\left(\dfrac{N}{K_{i+3}+N}\right),\left(\dfrac{K_i}{K_i+E}\right)\left(\dfrac{K_{i+3}}{K_{i+3}+N}\right), \\
    & \left(\dfrac{K_i}{K_i+E}\right)\left(\dfrac{S}{K_{i+3}+S}\right), \left(\dfrac{K_i}{K_i+E}\right)\left(\dfrac{K_{i+3}}{K_{i+3}+S}\right), \left(\dfrac{N}{K_i+N}\right)\left(\dfrac{S}{K_{i+3}+S}\right), \\
    & \left(\dfrac{N}{K_i+N}\right)\left(\dfrac{K_{i+3}}{K_{i+3}+S}\right), \left(\dfrac{K_i}{K_i+N}\right)\left(\dfrac{S}{K_{i+3}+S}\right), \left(\dfrac{K_i}{K_i+N}\right)\left(\dfrac{K_{i+3}}{K_{i+3}+S}\right)\Biggl\}\\
\end{align*}

As $F_i(E,N,S)$ has 24 possible production terms, and there are three production terms in Eqs. (\ref{agar_model1})-(\ref{agar_model2}). As a result, our total family of regulatory networks is comprised of 24$^3$ = 13,824 models. \\

To model the knock-out experiments, we removed all production terms from $F_i(E,N,S)$ that involved the protein associated with the gene that was knocked out. We model the \textit{eud-1} knock-out perturbation using the following family of regulatory network models

    \begin{align}
    \text{[nhr-40 mRNA]:} ~\dfrac{dM_n}{dt} &= \alpha_2+k_{t_{n}}G_2(E,N,K_2,K_5)-k_{-m} M_n, \label{eko_eq1} \\ 
    \text{[NHR-40 protein]:} ~\dfrac{dN}{dt} &= k_{t} M_n - k_{-p}N, \\
    \text{[sult-1 mRNA]:} ~\dfrac{dM_s}{dt} &= \alpha_3 + k_{t_{s}}G_3(E,N,K_3,K_6) -k_{-m} M_s, \\
    \text{[SULT-1 protein]:} ~\dfrac{dS}{dt} &= k_{t} M_s  - k_{-p}S. \label{eko_eq2} \\ 
\end{align}

where $G_i(N,S)$ for $i=2,3$ is defined as

\begin{align*}
    G_i(N,S) &=\\
    &\Biggl\{\left(\dfrac{N}{K_i+N}\right),\left(\dfrac{K_i}{K_i+N}\right),\left(\dfrac{S}{K_i+S}\right),\left(\dfrac{K_i}{K_i+S}\right),\\
    &\left(\dfrac{N}{K_i+N}\right)^2,\left(\dfrac{K_i}{K_i+N}\right)^2,\left(\dfrac{S}{K_i+S}\right)^2,\left(\dfrac{K_i}{K_i+S}\right)^2, \\
     &\left(\dfrac{N}{K_i+N}\right)\left(\dfrac{S}{K_{i+3}+S}\right), \left(\dfrac{N}{K_i+N}\right)\left(\dfrac{K_{i+3}}{K_{i+3}+S}\right), \\
     &\left(\dfrac{K_i}{K_i+N}\right)\left(\dfrac{S}{K_{i+3}+S}\right), \left(\dfrac{K_i}{K_i+N}\right)\left(\dfrac{K_{i+3}}{K_{i+3}+S}\right)\Biggl\}\\
\end{align*}

As $G_i(E,N,S) \subset F_i(E,N,S)$ and has 12 possible production terms, and there are two production terms in the Eqs. (\ref{eko_eq1})-(\ref{eko_eq2}). As a result, our total family of regulatory networks is comprised of 12$^2$ = 144 models. \\

Similarly, we model the \textit{sult-1} knock-out perturbation using the following family of regulatory network models

\begin{align}
   \text{[eud-1 mRNA]:} ~\dfrac{dM_e}{dt} &= \alpha_1 + k_{t_{e}}H_1(N,S,K_1,K_4)-k_{-m} M_e, \label{sko_eq1}\\
    \text{[EUD-1 protein]:} ~\dfrac{dE}{dt} &= k_{t} M_e - k_{-p}E, \\
    \text{[nhr-40 mRNA]:} ~\dfrac{dM_n}{dt} &= \alpha_2+k_{t_{n}}H_2(N,S,K_3,K_6)-k_{-m} M_n, \\ 
    \text{[NHR-40 protein]:} ~\dfrac{dN}{dt} &= k_{t} M_n - k_{-p}N. \label{sko_eq2}
\end{align}

where $H_i(E,N)$ for $i=1,2$ is defined as

\begin{align*}
    H_i(E,N,S) &=\\
    &\Biggl\{\left(\dfrac{E}{K_i+E}\right), \left(\dfrac{K_i}{K_i+E}\right),\left(\dfrac{N}{K_i+N}\right),\left(\dfrac{K_i}{K_i+N}\right),\\
    &\left(\dfrac{E}{K_i+E}\right)^2, \left(\dfrac{K_i}{K_i+E}\right)^2,\left(\dfrac{N}{K_i+N}\right)^2,\left(\dfrac{K_i}{K_i+N}\right)^2, \\
    &\left(\dfrac{E}{K_i+E}\right)\left(\dfrac{N}{K_{i+3}+N}\right), \left(\dfrac{E}{K_i+E}\right)\left(\dfrac{K_{i+3}}{K_{i+3}+N}\right), \\ 
    &\left(\dfrac{K_i}{K_i+E}\right)\left(\dfrac{N}{K_{i+3}+N}\right),\left(\dfrac{K_i}{K_i+E}\right)\left(\dfrac{K_{i+3}}{K_{i+3}+N}\right)\Biggl\}\\
\end{align*}

$H_i(E,N,S) \subset F_i(E,N,S)$ and has 12 possible production terms, and there are two production terms in the Eqs. (\ref{sko_eq1})-(\ref{sko_eq2}). As a result, our total family of regulatory networks is comprised of 12$^2$ = 144 models. \\

However, these models are not structurally identifiable. We address structural identifiability testing and removal of a scaling symmetry in Section \ref{structural} before fitting these models to the experimental data. 

\subsection{Structural identifiability testing}
\label{structural}
As shown in Section \ref{results}, the model library contains a scaling symmetry (see Eqs. (\ref{eqni1})-(\ref{eqni2})). This scaling symmetry impacts Eqs. (\ref{agar_model1})-(\ref{agar_model2}), Eqs. (\ref{eko_eq1})-(\ref{eko_eq2}), Eqs. (\ref{sko_eq1})-(\ref{sko_eq2}). We identified the scaling symmetry and found a reparameterization of the tested model using STRIKE-GOLDD \cite{massonis2020finding}. To eliminate the scaling symmetry, we set $\xi=\dfrac{1}{k_t}$, which rescales the protein states and the half-max parameters by the translation rate, which must be included in the model and can never be zero. Therefore, we scaled the proteins and half-max parameters by the translation rate. By inspection, we concluded that the scaling symmetry would impact the entire library, given the conserved structure of the equations. 

% We felt comfortable with this reparameterization, as it uses the translation rate to break the symmetry, which must be in the model structure and can never go to zero. Other alternative reparameterization options existed. In particular, the symmetry could also be broken by scaling by a half-max parameter, but there is no guarantee that any half-max parameter must be in the model. Therefore, we scaled the by the translation rate. By inspection, we concluded that the scaling symmetry would impact the entire library, given the conserved structure of the equations. 

To ensure the reparameterized models were structurally identifiable, we attempted to screen the entire family using an independent algorithm \cite{structidjl}. We screened the identifiability of all kinetic parameters and initial conditions for the entire family of ODE models (including both the wild-type models and the knock-out models) using the same algorithm, and found that thousands of models were indeed structurally identifiable, but this approach was computationally infeasible for many of the highly nonlinear models. Many highly non-linear models did not finish in 24 hours on an HPC cluster, so we cannot conclude if they are structurally identifiable or not. It is known that some model structures can cause long algorithm run-times \cite{structidjl}. Of the thousands of models that finished within 24 hours, we found that the kinetic parameters were identifiable, and the initial conditions of the scaled protein states were identifiable if they appeared in at least one of the mRNA production terms in the model.  It is also important to note that the family of ODE models could contain a more complex scaling or transformation symmetry that is impacting structural identifiability that we failed to detect, even though we used multiple algorithms to assess structural identifiability \cite{massonis2020finding,structidjl}.

% and found that they were. We also used the same algorithm to test for identifiability of the initial conditions associated with the candidate model (experimental setting) \cite{structidjl}. We found that the initial conditions for the mRNA states were identifiable, as they are measured states of the system, and the intial conditions of the scaled protein states were identifiable if they appeared in at least one of the mRNA productions terms in the model. It should be noted that this finding is based on testing a limited number of models. We screened the identifiability of all kinetic parameters and initial conditions for the entire nested family of ODE models (including both the wild-type models and the knock-out models) using the same algorithm, and found that thousands of models were indeed structurally identifiable, but this approach was computationally infeasible for many of the highly nonlinear models. Many highly non-linear models did not finish in 24 hours on an HPC cluster. 

% It is also important to note that the nested family of ODE models could contain a more complex scaling or transformation symmetry that is impacting structural identifiability that we failed to detect, even though we used multiple algorithms to assess structural identifiability \cite{structidjl}. 

\subsection{Parameter estimation}
\label{estimation}
We performed parameter estimation across the model structures for each experimental data set using the L-BFGS-B algorithm initialized from 1,000 places in parameter space using Latin-Hypercube sampling(see Sections \ref{models} and \ref{structural}) \cite{zhu1997algorithm}. As we did not have good prior estimates, we chose very wide lower and upper bounds for unscaled parameters, between 1x10$^{-5}$ and 1x10$^4$+1, respectively. For parameters scaled by the translation rate (see Section \ref{structural}), we allowed parameters to range from 1x10$^{-9}$ and 1x10$^8$+1. The Latin-Hypercube samples were logarithmically rescaled between 1x10$^{-4}$ and 1x10$^4$ and 1x10$^{-8}$ and 1x10$^8$ for unscaled and scaled parameters, respectively. Smaller parameter ranges cause some parameters to hit the upper or lower bounds we set. Tuning optimization hyperparameters, such as the upper and lower bounds, for individual model structures was infeasible due to the large number of models. We assumed the data followed a multiplicative error model \footnote{Specifically, we assume $\bar{y} =(1+\epsilon)\bar{x}$, where $\epsilon \sim \mathcal{N}(0, \sigma^2)$ and $\bar{y}$ is the vector of observed data.} and minimized the associated negative log-likelihood function across all experiments. The negative log-likelihood function has the form 
$$ -l(\boldsymbol{\theta},\sigma^2|y_{1\dots n}) = n \text{log}(\sigma) + \sum_{i=1}^{n} \text{log}(x_i(\boldsymbol{\theta})) + \dfrac{1}{2 \sigma^2} \sum_{i=1}^n \left(\dfrac{y_i -x_i(\boldsymbol{\theta})}{x_i(\boldsymbol{\theta})}\right)^2.$$
We performed this computation using the Quest High-Performance Computing Cluster at Northwestern University, then defined model sets based on the modeling results for each experiment (see Section \ref{sets}). 

To generate the tens of thousands of individual loss function and ODE model files, we used the templating language Jinja2 (\url{https://pypi.org/project/Jinja2/}). Numerical integration of ODEs was done using \texttt{OrdinaryDiffEq.jl}, a component package of \texttt{DifferentialEquations.jl} \cite{rackauckas2017differentialequations}. We used the \texttt{Optimization.jl} implementation of L-BFGS-B \cite{vaibhav_kumar_dixit_2023_7738525} and \texttt{LatinHypercubeSampling.jl} implementation of Latin-Hypercube Sampling \cite{urquhart_surrogate-based_2020}. Other packages used during our study include \texttt{Random.jl}, \texttt{LinearAlgebra.jl}, \texttt{DataFrames.jl}, \texttt{CSV.jl},  \texttt{SciMLSensitivity.jl} \cite{rackauckas2020universal}, \texttt{ForwardDiff.jl} \cite{RevelsLubinPapamarkou2016}, and \texttt{ModelingToolkit.jl} \cite{ma2021modelingtoolkit}.

\subsection{Model sets calculation}
\label{sets}

After fitting the models of interest (see Sections \ref{data}, \ref{models}, \ref{structural}, \ref{estimation}), we built model sets, or models with shared regulatory network features that generate acceptable fits to the data. It was feasible to build the model sets associated with each of the knock-out experiments by hand, given the relatively few models involved and the clear discrete jump in the cost function value (see Figure \ref{fig:5} A, middle and right). Building the model set associated with the wild-type experimental data was more involved.

To find the model set associated with the wild-type experiment, we first binarized 21 model features of the 13,824 model structures we tested. We then used a decision tree to predict if the models gave an acceptable or unacceptable fit to the data based on the model features (see Figure \ref{fig:decision_tree}). We considered the models with the 1000 lowest cost function values to be acceptable fits (Figure \ref{fig:4} A, pink) and all other model fits to be unacceptable (Figure \ref{fig:4} A, gray) based on the behavior of the cost function value curve in Figure \ref{fig:4}A. After 1000 models, the cost function value curve plateaus, then increases dramatically (see Figure \ref{fig:4}A). We used K-Fold cross-validation and found an average accuracy of 94.8\%. Note that this is higher than a null-hypothesis of all models do not produce an acceptable fit (92.8\% accuracy). 

We explored up-sampling the minority class and found that it led to a marginal improvement in accuracy, but also led to overfitting and did not qualitatively change the results. To decide on the depth of the tree, we swept over the depth parameter of the decision tree, and found that accuracy was maximized when the maximum depth was between five and seven. We ran the analysis using both Gini index and entropy as criteria and found the result did not change. For simplicity, we set the maximum depth to five. Using this approach, we were able to detect a structural signal of the combination of model features that generate an acceptable fit to the data, but were not able to correctly classify all models. This is unsurprising, as the cost function value curve exhibits a rounded plateau (Figure \ref{fig:4}A). 

\section{Author Contributions}
C.E.F., M.S.W., and N.M.M. designed the research; C.E.F. and N.M.M. oversaw the research;  C.E.F., A.M., and V.A. performed research; C.E.F., S.R., A.M., and M.S.W. wrote the paper; All authors reviewed and edited the paper. 

\section{Competing Interests}
The authors declare no competing interest.

\section{Supplemental Figure}

\begin{figure}[H]
    \begin{center}
    \includegraphics[height=0.55\textwidth,angle=90]{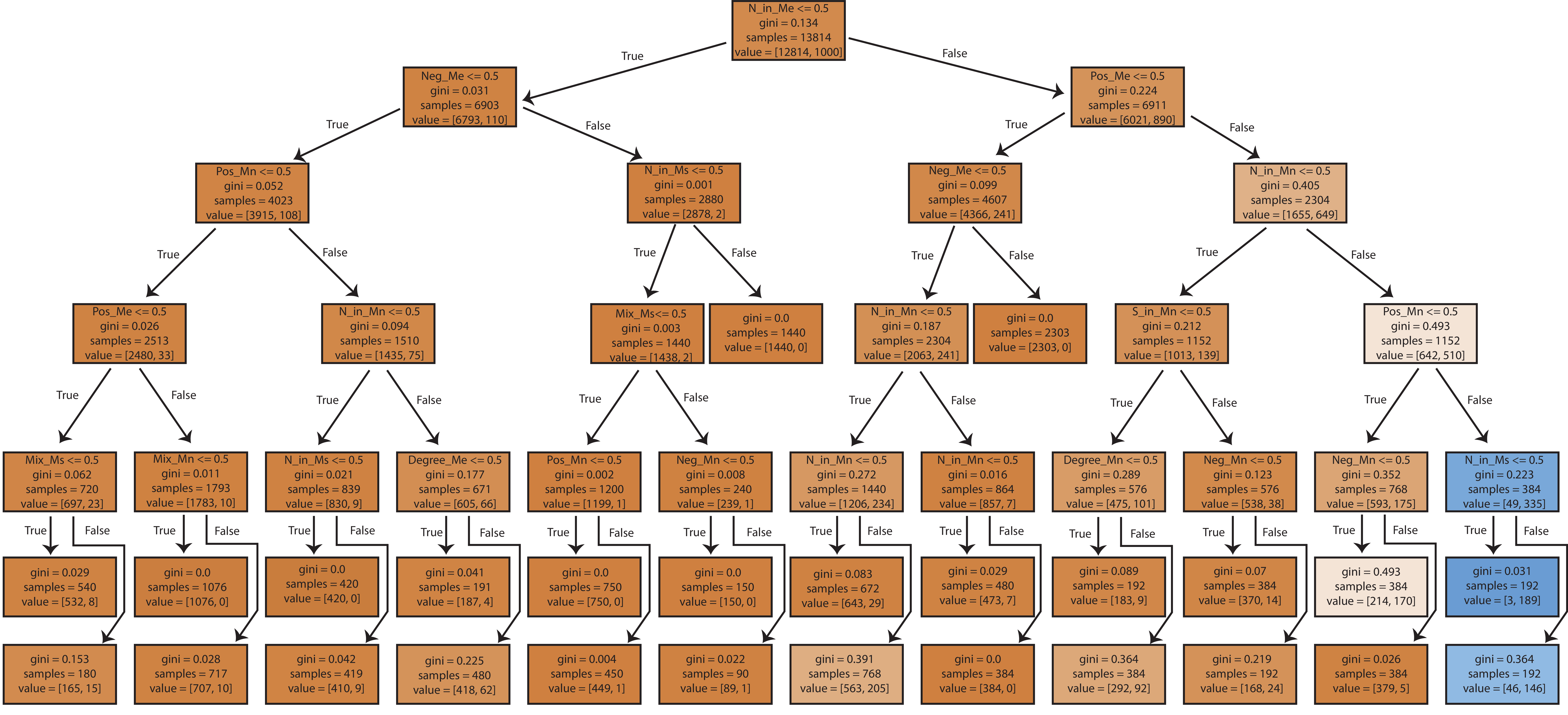}
    \end{center}
    \caption{(Caption continued on next page.)}
    \label{fig:decision_tree}
\end{figure}

\newpage

\begin{figure}[t]
  \contcaption{We used a decision tree to classify models that produced an acceptable fit to the wild type experiment (Figure \ref{fig:4} A, pink) from those that produced an unacceptable fit (Figure \ref{fig:4} A, gray) based on 21 distinct structural features of the 13,814 models (note: optimization did not finish within the 4 hour wall time for 10 models and are excluded from the analysis). The binary structural features of the model we used during the classification included are E\_in\_Me, N\_in\_Me, S\_in\_Me, Pos\_Me, Neg\_Me, Mix\_Me, Degree\_Me, E\_in\_Mn, N\_in\_Mn, S\_in\_Mn, Pos\_Mn, Neg\_Mn, Mix\_Mn, Degree\_Mn, E\_in\_Ms, N\_in\_Ms, S\_in\_Ms, Pos\_Ms, Neg\_Ms, Mix\_Ms, Degree\_Ms which represent $\hat{E}$ regulates \textit{eud-1} mRNA production term, $\hat{N}$ regulates \textit{eud-1} mRNA production term, $\hat{S}$ regulates \textit{eud-1} mRNA production term, \textit{eud-1} mRNA is solely positively regulated, \textit{eud-1} mRNA is solely negative regulated, \textit{eud-1} mRNA is modeled by a production term involving only one protein with an exponent of one, $\hat{E}$ regulates \textit{nhr-40} mRNA production term, $\hat{N}$ regulates \textit{nhr-40} mRNA production term, $\hat{S}$ regulates \textit{nhr-40} mRNA production term, \textit{nhr-40} mRNA is solely positively regulated, \textit{nhr-40} mRNA is solely negative regulated, \textit{nhr-40} mRNA is both positively and negatively regulated, \textit{nhr-40} mRNA is modeled by a production term involving only one protein with an exponent of one, $\hat{E}$ regulates \textit{sult-1} mRNA production term, $\hat{N}$ regulates \textit{sult-1} mRNA production term, $\hat{S}$ regulates \textit{sult-1} mRNA production term, \textit{sult-1} mRNA is solely positively regulated, \textit{sult-1} mRNA is solely negative regulated, \textit{sult-1} mRNA is both positively and negatively regulated, \textit{sult-1} mRNA is modeled by a production term involving only one protein with an exponent of one, respectively. The value 0 corresponds to FALSE and the value 1 corresponds to TRUE for each of the binary structural features of the model. `Gini', `samples', and `value' are the Gini impurity value (a measure of split quality), the number of model structures that reach each node of the tree, and how samples are split by class (the first index corresponds to the unacceptable class and the second index corresponds to the acceptable class), respectively.}% Continued caption
\end{figure}

\section{Funding}

  C. E. F. is supported in part by The James S. McDonnell Foundation Postdoctoral Fellowship Award in Complex Systems (\url{https://doi.org/10.37717/2020-1591}) and by the NSF-Simons Center for Quantitative Biology at Northwestern University (NSF: 1764421 and Simons Foundation/SFARI 597491-RWC). N. M. M. was supported by the U.S. Department of Energy, Office of Science, Office of Advanced Scientific Computing Research, under Award Number DESC0024253. V.A. was supported as an undergraduate research assistant by the U.S. Department of Energy, Office of Science, Office of Advanced Scientific Computing Research, under Award Number DESC0024253. A.M. was supported as a Summer Undergraduate Research Program Participant by NSF (DMS-2235451) and Simons Foundation (MPS-NITMB-00005320). M.S.W. and S.R. were funded by the National Institutes of Health, National Institute of General Medical Sciences grant R35GM150720-01. 

  This research was supported in part by grants from the NSF (DMS-2235451) and Simons Foundation (MP-TMPS-00005320) to the NSF-Simons National Institute for Theory and Mathematics in Biology (NITMB). This research was supported in part through the computational resources and staff contributions provided for the Quest high performance computing facility at Northwestern University which is jointly supported by the Office of the Provost, the Office for Research, and Northwestern University Information Technology.

\section{Acknowledgements}

C.E.F. thanks Scotty Coughlin and Matthew Gorby of Research Computing Services at Northwestern University for their help in HPC design and the scientific community of the National Institute for Theory and Mathematics in Biology for helpful comments that improved the work. 

\bibliographystyle{ieeetr}
\bibliography{mouthform.bib}

\end{document}